# Comment: Bayesian Checking of the Second Levels of Hierarchical Models

**Andrew Gelman**

Bayarri and Castellanos (BC) have written an interesting paper discussing two forms of posterior model check, one based on cross-validation and one based on replication of new groups in a hierarchical model. We think both these checks are good ideas and can become even more effective when understood in the context of posterior predictive checking. For the purpose of discussion, however, it is most interesting to focus on the areas where we disagree with BC:

1. We have a different view of model checking. Rather than setting the goal of having a fixed probability of rejecting a true model and a high probability of rejecting a false model, we recognize ahead of time that our model is wrong and view model checking as a way to explore and understand differences between model and data.
2. BC focus on $p$-values and scalar test statistics. We favor graphical summaries of multivariate test summaries.
3. For BC, it is important that $p$-values have a uniform distribution (i.e., that they be $u$-values, in our terminology) under the assumption that the null hypothesis is true. For us, it is important that $p$-values be interpretable as posterior probabilities comparing replicated to observed data.
4. BC recommend an "empirical Bayes prior $p$-value" as being better than the posterior predictive $p$-value. In fact, their empirical Bayes prior $p$-value is an approximation to a posterior predictive $p$-value which was recommended for hierarchical models in Gelman, Meng and Stern (1996). BC miss this connection by not seeing the full generality of posterior predictive checking.

In our discussion, we go through each of the above points in turn and conclude with a comment on the potential importance of theoretical work such as BC's on the future development of predictive model checking.

## 1. THE GOAL OF MODEL CHECKING: REJECTING FALSE MODELS, OR UNDERSTANDING WAYS IN WHICH THE MODEL DOES NOT FIT DATA

All models are wrong, and the purpose of model checking (as we see it) is not to reject a model but rather to understand the ways in which it does not fit the data. From a Bayesian point of view, the posterior distribution is what is being used to summarize inferences, so this is what we want to check. The key questions then become: (a) what aspects of the model should be checked; (b) what replications should we compare the data to; (c) how to visualize the model checks, which are typically highly multi-dimensional; (d) what to make of the results?

In a wide-ranging discussion of a range of different methods for Bayesian model checking, BC focus on the above question (d): in particular, how can Bayesian hypothesis testing be set up so that the resulting $p$-values can used as a model-rejection rule with specified Type I errors? This question is sometimes framed as a desire for calibration in $p$-values, but ultimately the desire for calibration is most clearly interpretable within a model-rejection framework. For example, BC write that some methods "can result in a severe conservatism incapable of detecting clearly inappropriate models." But it is not at all clear that, just because a model is wrong, that it is "inappropriate." If a model predicts replicated data that are just like the observed data in important ways, it may very well be appropriate for these purposes. Recall that we have already agreed that our models are wrong; we would like to measure appropriateness in a direct way, rather than set a rule that even a true model must be declared "inappropriate" 5% of the time. For example, in the model considered by BC, we do not see the rationale for their testing the hypothesis $\mu = \mu_0$; we would rather just perform Bayesian inference for $\mu$.

*Andrew Gelman is Professor, Departments of Statistics and Political Science, Columbia University, New York, New York 10027, USA e-mail: gelman@stat.columbia.edu*

Our concerns are thus a bit different from those of BC: we are less concerned about the properties of our procedures in the (relatively uninteresting) case that the model is true, and more interested in having the ability to address the misfit of model to data in direct terms. One reason, perhaps, of the popularity of our posterior predictive approach, in addition to its Bayesian flavor and ease of implementation, is the flexibility that allows us to consider complicated test summaries—including plots of the entire data set, as well as combinations of data and parameters and combinations of observed and missing data—thus bringing the power of exploratory data analysis to the checking of Bayesian models, and conversely bringing the power of Bayesian inference to exploratory data analysis.

Some of the difference in focus can be seen by looking at the graphs in BC—histograms of the null distributions of $p$-values, curves of predictive densities of unidimensional test summaries, and a single plot of raw data (but with no comparison plots of replicated data)— and comparing to the graphs in Gelman (2004) and Gelman et al. (2005), which show various plots of time series and other multidimensional test summaries.

## 2. THE STEPS OF BAYESIAN MODEL CHECKING

BC begin their paper with a useful characterization of any checking method as having a diagnostic statistic, a distribution for the statistic, and a way to measure conflict with the null distribution. Here we briefly explain how our own applied model checking fits into BC's three-step framework.

*Step* 1. BC consider a diagnostic statistic $T(x_{\text{obs}})$ that depends entirely on observed data. In a Bayesian framework, the diagnostic statistic, or test statistic, or discrepancy measure can also depend on parameters (Gelman, Meng and Stern, 1996) and on missing or latent data (Gelman et al., 2005). It can be helpful to look purely at observed data, but the expanded formulation can allow us to define test variables that more directly catch features of substantive interest.

*Step* 2. We compare the test variable to the predictive distribution of other data sets $y^{\text{rep}}$ that could have arisen from the same model. Formally introducing the replications $y^{\text{rep}}$ is an important step in the mathematical formulation of Bayesian testing because it makes explicit the joint model, $p(y, y^{\text{rep}}, \theta)$. (Bayarri and Castellanos use the notation $x$ for data, but we prefer y because we commonly work in the applied regression framework in which $y$ is modeled conditional on predictors, $x$.) Because we are doing Bayesian inference, we simply use the posterior distribution, $p(y^{\text{rep}}|y)$, which is also called the posterior predictive distribution because $y^{\text{rep}}$ can be viewed as predictions.

As discussed by Gelman, Meng and Stern (1996), the prior predictive distribution is also a posterior predictive distribution but with $y^{\text{rep}}$ defined as arising from new parameters, $\theta^{\text{rep}}$, drawn from the model. The choice of prior or posterior distribution—or, more generally, the choice of what is to be replicated in defining $y^{\text{rep}}$—depends on which aspects of the model are being checked. In many cases, the prior distribution is assigned based on convenience and so there is no particular interest in checking its fit to the data.

In the context of the paper at hand, which is explicitly concerned with checking the second level of a hierarchical model, it makes sense to use an intermediate replication, in which the hyperparameters $\eta$ are kept the same but the lower-level parameters $\theta$ are replicated—that is, resampled from the group-level model. In the notation of BC, the predictive distribution of interest would be $p(\theta^{\text{rep}}, x^{\text{rep}}, \eta|x)$, averaging over the posterior distribution $p(\eta|x)$. This is a slight departure from BC's recommendation to integrate $\theta$. (Actually, we prefer the term "average over" to "integrate out" since we perform our computations using simulation.) As we discuss in Section 4 below, it turns out this is very close to what BC call the empirical Bayes prior predictive check.

*Step* 3. For a one-dimensional test summary, the discrepancy between model and data can be summarized by a $p$-value or, often more usefully, by a predictive confidence interval. (For example, page 366 of Bayesian Data Analysis has an example from an analysis of elections in which 12.6% of the elections in the data switched parties, but in replicated data sets the 95% interval for the proportion of switches was [13.0%, 14.3%]. In this case, the model clearly did not fit this aspect of the data, but this difference of about one percentage point was not of practical significance.) For higher-dimensional test summaries, graphical summaries would be appropriate— up to and including plots of the entire data set, compared with plots of replicated data. There is some potential, we believe, to connect classes of models with classes of graphs to suggest natural and automatic displays of checks for many problems (Gelman, 2003, 2004).

As we have already noted, BC focus on $p$-values, which can be useful summaries but are no replacement for graphical comparisons of observed and replicated data that can reveal various aspects of model misfit. We emphasize that any of the methods discussed in the BC paper can be applied to graphical checks.

## 3. $p$-VALUES AND $u$-VALUES

Regarding the discussion in Section 3.5 of BC on $p$-values, we refer the reader to Section 2.3 of Gelman (2003), which distinguishes between Bayesian $p$-values—most simply, posterior probability statements of the form $\Pr(T(y^{\text{rep}}) > T(y)|y)$—and $u$-values—data summaries with a uniform null distribution. Classical $p$-values with pivotal test statistics are also $u$-values, but in the presence of uncertainty about parameters it is not generally possible for tests to have both properties at once. On the occasions that we do summarize test statistics using tail-area probabilities, we prefer the $p$-value because it can be directly interpreted as a statement, conditional on the model, about what might be expected in future replications. Here we disagree with BC, who describe the uniform null distribution as "a very desirable property, namely having the same interpretation across problems." It is perhaps a matter of taste whether to prefer a posterior summary with a direct probabilistic interpretation or a less-interpretable statistic that has a uniform distribution under the null model. We would certainly not call our $p$-values uninterpretable: for example, a $p$-value of 0.2 means clearly that, under the model, 20% of future data will be at least as extreme as the observed data. No calibration is necessary for this interpretation to be valid.

In any case, our point here is to distinguish between the two goals—a direct probability statement and a uniform null distribution—and to point out that, in general, you cannot have both, just as, in general, posterior means will not be unbiased estimates and posterior intervals will not have classical confidence coverage for all parameter values. Ultimately we will evaluate our Bayesian model-checking methods based on how well they help us understand differences between model and data, not based on theoretical coverage properties and not based on their rates of rejecting models which we know are false anyway.

## 4. THE "EMPIRICAL BAYES PRIOR $p$-VALUE"

BC's paper concludes with a statement that empirical Bayes prior $p$-values "have better properties [than posterior $p$-values] and are easier to compute." In fact, these EB-prior $p$-values are very close to posterior $p$-values, replicating $\theta$ but leaving the hyperparameters ($\eta$, in BC's notation) fixed, a strategy which Gelman, Meng and Stern (1996) recommend for hierarchical models (Figure 1c on page 739 of that paper). The only difference between the EB-prior distribution and this posterior predictive distribution is that the former uses point estimates of the hyperparameters, which cannot in general be a good idea (consider, e.g., settings where no good point estimates exist, such as the 8-schools example from Chapter 5 of Bayesian Data Analysis). We suspect the good performance of the EB-prior $p$-values comes from the appropriate choice of replication for testing the second level of a hierarchical model—the same hyperparameters but new groups—not from the use of point estimates.

To put it another way, take BC's "empirical Bayes" method, average over the hyperparameters so that it becomes "hierarchical Bayes" (as is appropriate given the other parts of the paper), and you get a posterior predictive check. We suppose that BC did not notice this because of their assumption that in posterior predictive checking, all parameters had to be kept the same in replications (as in Figure 1a on page 739 of Gelman, Meng and Stern, 1996). In fact, the flexibility of predictive checking allows different aspects of the data and parameter vectors to be preserved in replications, and for the particular goal of BC's paper, it makes sense to replicate the parameters $\theta$ (as BC ended up discovering in their simulations). Sinharay and Stern (2003) discuss these issues further in the context of the hierarchical normal model.

## 5. LOOKING FORWARD

As indicated by the plethora of methods discussed by BC, there are many ways of combining ideas of replication and cross-validation. A parallel situation arises in the literature of the bootstrap (Efron and Tibshirani, 1993), with parametric bootstraps, nonparametric bootstraps, and special methods for spatial and time-series data. A lot more work needs to be done. In particular, although we do find the posterior predictive framework useful, we recognize that there is something particularly compelling about external validation and cross-validation. At the theoretical level, there is an opening to incorporate validation into hierarchical modeling with the possibilities of different levels of cross-validation for individuals and groups (e.g., fivefold cross-validation of groups and tenfold cross-validation of observations within groups). More practical concerns include decisions about how to set up the tuning parameters for cross-validation and, when comparisons are made graphically, how to visualize the many replicated data sets. BC's partial posterior predictive distribution could be an excellent way to unify this area.

The BC paper focuses on $p$-values, but if our own experience is any guide, we expect the most useful work to focus on graphical explorations of realized and replicated data. We focused on $p$-values in our 1996 paper, but in the years since, we have found graphical checks to be more helpful, with numerical summaries and $p$-values coming in at the end to give some structure to our visual judgments. The theoretical structure used by BC, of looking at null distributions of $p$-values, could become helpful here, and also for concerns of multiple comparisons.

## ACKNOWLEDGMENTS

We thank Hal Stern for helpful comments and the National Science Foundation and National Institutes of Health for financial support.